\documentclass[aps,pra,twocolumn,superscriptaddress]{revtex4}

\usepackage{amsmath,amssymb,amsthm,color,mathrsfs}
\usepackage{amsbsy,mathtools}
\usepackage{hyperref}
\usepackage{graphicx}
\usepackage{bm}

\newcommand{\p}{\partial}

\newcommand{\ex}{\mathbf{\hat{e}}_1}
\newcommand{\ey}{\mathbf{\hat{e}}_2}
\newcommand{\ez}{\mathbf{\hat{e}}_3}
\newcommand{\emn}{\epsilon_{\mu\nu}}

\newcommand{\DM}{D}
\newcommand{\dm}{\lambda}
\newcommand{\Anisotropy}{K}
\newcommand{\anisotropy}{\kappa}

\newcommand{\Energy}{W}
\newcommand{\Eex}{W_{\rm ex}}

\newcommand{\xmagn}{X}

\newcommand{\skyrmion}{q}
\newcommand{\Skyrmion}{Q}

\newcommand{\ldm}{\ell_{\rm D}}
\newcommand{\storque}{\beta}
\newcommand{\storquedw}{u}

\begin{document}

\title{Skyrmion dynamics in chiral ferromagnets under spin-transfer torque}
\author{Stavros Komineas}
\affiliation{Department of Mathematics and Applied Mathematics, University of Crete, 71003 Heraklion, Crete, Greece}
\author{Nikos Papanicolaou}
\affiliation{Department of Physics, University of Crete, 71003 Heraklion, Crete, Greece}
\begin{abstract}
We study the dynamics of skyrmions under spin-transfer torque 
in Dzyaloshinskii-Moriya materials with easy-axis anisotropy.
In particular, we study the motion of a topological skyrmion with skyrmion number $\Skyrmion=1$ and a
non-topological skyrmionium with $\Skyrmion=0$
using their linear momentum, virial relations, and numerical simulations.
The non-topological $\Skyrmion=0$ skyrmionium is accelerated in the direction of the current flow 
and it either reaches a steady state with constant velocity, or it is elongated to infinity.
The steady-state velocity is given by a balance between current and dissipation and has an upper limit.
In contrast, the topological $\Skyrmion=1$ skyrmion converges to a steady-state with constant velocity
at an angle to the current flow.
When the spin current stops 
the $\Skyrmion=1$ skyrmion is spontaneously pinned
whereas the $\Skyrmion=0$ skyrmionium continues propagation.
Exact solutions for the propagating skyrmionium are identified as solutions of equations
given numerically in a previous work. 
Further exact results for propagating skyrmions are given in the case of the pure exchange model.
The traveling solutions provide arguments that a spin polarized current will cause
rigid motion of a skyrmion or a skyrmionium.
\end{abstract}

\pacs{75.76.+j,	%Spin transport effects
75.78.-n,   % Magnetization dynamics
75.70.Kw   % Domain structure (including magnetic bubbles and vortices)
}

\maketitle

%%%%%%%%%%%%%%%%%%%%%%%%%%%%%%%%%%%%
\section{Introduction}
\label{sec:intro}

Soliton structures are found in ferromagnets and they can be considered as an encoding of magnetic information
which is robust both under temperature and external probes.
Stable topological solitons with the structure of a skyrmion 
had been predicted in the presence of the Dzyaloshinskii-Moriya (DM) interaction
\cite{BogdanovYablonskii_JETP1989,BogdanovHubert_JMMM1994}
and they were observed in recent years as
isolated structures \cite{RommingHanneken_Science2013,RommingKubetzka_PRL2015} or forming lattices \cite{MuhlbauerBinz_Science2009,YuOnose_Nature2010,YuKanazawa_NatMat2011}.
In the presence of easy-axis anisotropy there are topological skyrmions with skyrmion number $\Skyrmion=1$  
as well as non-topological $\Skyrmion=0$ solitons ($2\pi$ vortices) \cite{BogdanovHubert_JMMM1999,LeonovRoessler_EPJ2013}.

Skyrmions could be the stable and robust entities that are needed for the technology of
recording and transferring information, currently mainly obtained in magnetic media using domain walls
\cite{AllwoodXiong_Science2005}.
The propagation of magnetic information is done most conveniently by the injection
of electrical spin-polarized current.
Single skyrmions and skyrmion lattices can be set in motion suggesting a promising technique 
for the manipulation of magnetic information
\cite{EverschorGarst_PRB2012,FertCros_NatNano2013,SampaioCros_NatNano2013,NagaosaTokura_NatNano2013,IwasakiMochizuki_NatComms2013,IwasakiMochizuki_NatNano2013,SchutteIwasaki_PRB2014}.
Propagation of skyrmions by spin current or the related spin-Hall effect may be a promising strategy for the implementation of racetrack memories \cite{TomaselloMartinez_SciRep2014}.

The existence of two species of skyrmions ($\Skyrmion = 0$ and $\Skyrmion \neq 0$)
has allowed theoretical predictions for dramatically different dynamical behaviors
\cite{KomineasPapanicolaou_PRB2015}.
We show here that spin torque accelerates a $\Skyrmion = 0$ skyrmionium and we describe the process theoretically, 
while the study is complemented by numerical simulations.
The skyrmionium may reach a steady state or it may absorb energy from the current and expand without limit.
A skyrmionium propagating even when the external probe is switched off can be obtained.
The situation is contrasted to the more well-studied case of a $\Skyrmion = 1$ skyrmion under spin torque.

The outline of the paper is the following.
Sec.~\ref{sec:model} gives a description of the Landau-Lifshitz model including damping and spin-transfer torques
and provides the main theoretical tools.
Sec.~\ref{sec:traveling_skyrmionium} is on the dynamics of a $\Skyrmion=0$ skyrmionium and
Sec.~\ref{sec:traveling_skyrmion} is on the dynamics of a $\Skyrmion=1$ skyrmion.
Sec.~\ref{sec:conclusions} contains our concluding remarks.
An Appendix gives analytical results on the pure exchange model.

%%%%%%%%%%%%%%%%%%%%%%%%%%%%%%%%%%%%
\section{Magnetization dynamics under spin-transfer torque}
\label{sec:model}

We assume a thin ferromagnetic film with a Dzyaloshinksii-Moriya (DM) interaction. 
Let $\bm{M}(x,y,t)$ be the magnetization vector with $M_s$ the saturation magnetization and
define  the normalised magnetization $\bm{m} = \bm{M}/M_s$, so that $\bm{m}^2=1$.
The conservative Landau-Lifshitz (LL) equation for the statics and dynamics of the magnetization is
\begin{equation}  \label{eq:LL}
\p_t \bm{m} = -\bm{m}\times\bm{f}
\end{equation}
and is valid in the absence of damping and external probes.
We consider an effective field $\bm{f}$ which includes an exchange interaction with constant $A$,
an easy-axis anisotropy perpendicular to the $(x_1,x_2)$-plane of the film with constant $\Anisotropy$,
and a DM interaction with constant $\DM$ \cite{BogdanovHubert_JMMM1994}.
If the energy is $\Energy$ then the effective field $\bm{f}=-\delta\Energy/\delta\bm{m}$ is
\begin{align}  \label{eq:effective_field}
\mathbf{f} & = \Delta\mathbf{m} + \anisotropy\, m_3 \ez   \\
  & - 2\dm\,\left[ \p_2 m_3\, \ex - \p_1 m_3\,\ey + (\p_1 m_2 - \p_2 m_1)\,\ez \right]  \notag
\end{align}
where we have used $\ldm = 2A/|D|$ as the unit of length.
The parameter
\begin{equation}   \label{eq:anisotropy}
\anisotropy \equiv \frac{\Anisotropy}{\Anisotropy_0},\qquad \Anisotropy_0 = \frac{D^2}{4A}
\end{equation}
is the rationalized (dimensionless) anisotropy constant
and $\dm = D/|D| = \pm 1$ will be referred to as the chirality.
We choose chirality $\dm=1$ in all of our numerical calculations,
while $\anisotropy$ is taken to be positive (easy-axis anisotropy).
We have not included the demagnetizing field in Eq.~\eqref{eq:effective_field} 
because it does not affect skyrmion configurations in a qualitatively significant way \cite{BegChernyshenko_arXiv2014};
it introduces a dependence of the skyrmion size on the film thickness \cite{KiselevBogdanov_JPD2011}.
The time variable $t$ in Eq.~\eqref{eq:llg_stt_ip} is measured in units of $\tau_0=2A M_s/(\gamma \DM^2)$ where
$\gamma$ is the gyromagnetic ratio.

When a spin-polarized current is flowing in the plane of the film, say, in the $x_1$ direction, 
and we also include damping effects, 
the magnetization obeys the Landau-Lifshitz-Gilbert equation with two additional terms for the spin-transfer torque
\cite{Slonczewski_JMMM1996,Berger_PRB1996,ZhangLi_PRL2004}:
\begin{equation}  \label{eq:llg_stt_ip}
(\p_t + \storquedw\, \p_1)\bm{m} = -\bm{m}\times\bm{f} + \bm{m}\times \left( \alpha\p_t + \storque u\, \p_1 \right) \bm{m}.
\end{equation}
The dissipation constant is $\alpha$ while $\storquedw$ is the effective spin velocity parallel to the spin current
and $\storque$ is the non-adiabatic spin-transfer torque parameter.

In the absence of spin torque, that is, for $\storquedw=0$ the ground state of the model is the {\it spiral state}
for sufficiently small anisotropy.
For $\anisotropy > \anisotropy_c = \pi^2/4\approx 2.4674$ the ground state is either of the two 
uniform ferromagnetic states $\bm{m}=(0,0,\pm 1)$.
In the latter case, skyrmions are excited states and they are classified by
the skyrmion number defined as
\begin{equation}  \label{eq:skyrmion_number}
\Skyrmion = \frac{1}{4\pi} \int \skyrmion\, d^2x,\qquad 
\skyrmion = \frac{1}{2}  \emn \mathbf{m}\cdot(\p_\nu\mathbf{m}\times \p_\mu\mathbf{m}),
\end{equation}
where $\skyrmion$ is called the {\it topological density}.
The skyrmion number $\Skyrmion$ is integer-valued ($\Skyrmion=0,\pm 1,\pm 2,\ldots$)
for all magnetic configurations such that $\mathbf{m}=(0,0,\pm 1)$ at spatial infinity.
For definiteness we will assume $\mathbf{m}=(0,0,1)$ in all our calculations.

\begin{figure}[t]
\begin{center}
\includegraphics[width=6truecm]{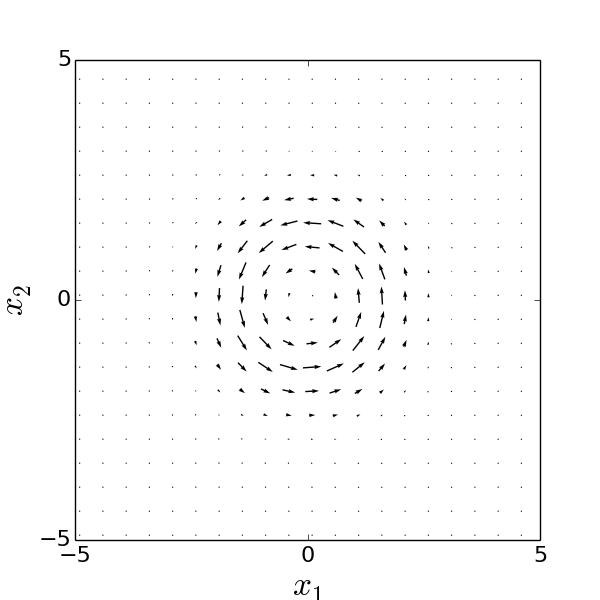}
\caption{The axially symmetric ($\Skyrmion=1$) skyrmion  represented through the projection
$(m_1,m_2)$ of the magnetization vector on the plane.
It is calculated as a static solution of Eq.~\eqref{eq:LL} for anisotropy $\anisotropy=3$.}
\label{fig:skyrmion_vecs}
\end{center}
\end{figure}

\begin{figure}[t]
\begin{center}
\includegraphics[width=6truecm]{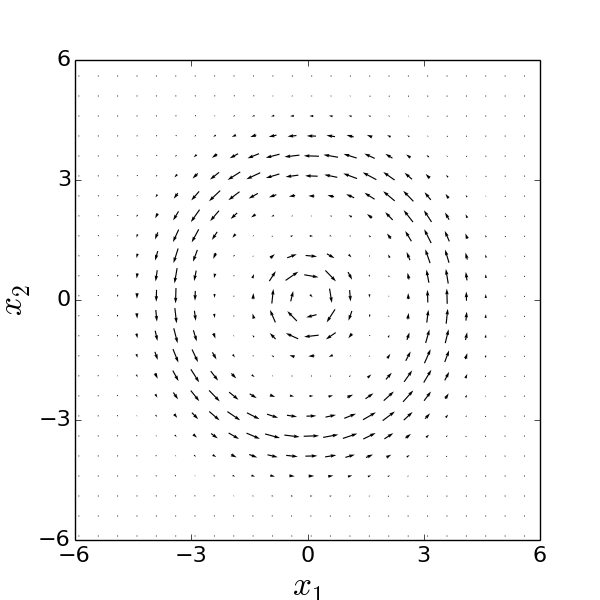}
\caption{The axially symmetric ($\Skyrmion=0$) skyrmionium represented through the projection
$(m_1,m_2)$ of the magnetization vector on the plane
It is calculated as a static solution of Eq.~\eqref{eq:LL} for anisotropy $\anisotropy=3$.}
\label{fig:skyrmionium_vecs}
\end{center}
\end{figure}

Axially symmetric skyrmion configurations
are conveniently described in terms of the standard spherical parametrisation given by
\begin{equation}
m_1 = \sin\Theta \cos\Phi,\quad m_2 = \sin\Theta \sin\Phi,\quad m_3 = \cos\Theta
\end{equation}
with the ansatz
\begin{equation}  \label{eq:axially_symmetric}
\Theta = \theta(\rho),\qquad \Phi = \phi + \pi/2,
\end{equation}
where $(\rho,\phi)$ are polar coordinates.
Solving Eq.~\eqref{eq:LL} with boundary conditions
$\theta(\rho=0)=\pi$ and $\theta(\rho\to\infty)=0$,
leads to a static skyrmion with $\Skyrmion=1$ shown in Fig.~\ref{fig:skyrmion_vecs}.
If the boundary conditions are $\theta(\rho=0)=2\pi, \theta(\rho\to\infty)=0$
a $\Skyrmion=0$ configuration is found \cite{BogdanovHubert_JMMM1999}
which has been called a ``skyrmionium'' \cite{FinazziSavoini_PRL2013,KomineasPapanicolaou_PRB2015}
and is shown in Fig.~\ref{fig:skyrmionium_vecs}.

%motion of the guiding centre or linear momentum
For the conservative LL Eq.~\eqref{eq:LL} it can be proved that the moments
of the topological density defined by
\begin{equation}  \label{eq:topological_moments}
I_\mu = \int x_\mu\skyrmion\, d^2x,\qquad \mu=1,2
\end{equation}
are conserved quantities \cite{PapanicolaouTomaras_NPB1991,KomineasPapanicolaou_PhysD1996}.
The conservation laws hold in the case of an infinite film.
For a magnetization field obeying an equation of the form $\p_t \bm{m}= -\bm{m}\times\bm{g}$ 
the time derivative of the topological density is
\begin{equation} \label{eq:vorticity_time_derivative}
\dot{\skyrmion} = -\emn\,\p_\mu (\bm{g}\cdot\p_\nu\bm{m})
\end{equation}
as can be found by a straightforward calculation.
For $\bm{g}=\bm{f}$ it can be shown that $\dot{I}_\mu=0$ and the ingredients of the proof for the specific $\bm{f}$
of Eq.~\eqref{eq:effective_field} are given in Ref.~\cite{KomineasPapanicolaou_PRB2015}.

Eq.~\eqref{eq:llg_stt_ip} including damping and spin torques
can be written in the form
\begin{align}
& \p_t \bm{m} = -\bm{m}\times\bm{g} \\
& \bm{g} = \frac{1}{1+\alpha^2} \left[ \bm{f} + \alpha\,\bm{m}\times\bm{f}
 - (\storque-\alpha) \storquedw\, \p_1\bm{m}
 - \alpha(\storque-\alpha) \storquedw\,\bm{m}\times\p_1\bm{m} \right]  \notag
\end{align}
and the time derivatives for $I_\mu$ can be found by using Eq.~\eqref{eq:vorticity_time_derivative}.
The moments $I_\mu$ are no longer conserved due to the damping and spin torque terms and we find
\begin{align}  \label{eq:virial1_stt_inplane}
(1+\alpha^2)\dot{I}_1 & = - (\storque-\alpha)\storquedw\, d_{12} + \alpha\, D_2 + (1+\alpha\storque) \storquedw\, (4\pi\Skyrmion)  \notag \\
(1+\alpha^2) \dot{I}_2 & = (\storque-\alpha)\storquedw\, d_{11} - \alpha\, D_1
\end{align}
where we have used the notation
\begin{equation}  \label{eq:d_munu}
\begin{split}
d_{\mu\nu} & = \int (\p_\mu\bm{m}\cdot\p_\nu\bm{m})\, d^2x,\qquad \qquad \mu,\nu=1,2  \\
D_\mu & = \int (\bm{m}\times \bm{f} ) \cdot\p_\mu\bm{m}\, d^2x.
\end{split}
\end{equation}
Eqs.~\eqref{eq:virial1_stt_inplane} give an explicit result since they may be applied for any
magnetic configuration.

% virial relations
Let us consider a skyrmion which is initially static within the conservative LL Eq.~\eqref{eq:LL}
and we suddenly apply an electrical current according to Eq.~\eqref{eq:llg_stt_ip}.
This will start to move and the overall motion is given by Eqs.~\eqref{eq:virial1_stt_inplane}.
As a next step we will assume that it will eventually reach a steady-state
with velocity $\bm{v}=(v_1, v_2)$.
We may then write the traveling wave ansatz for the magnetization
\begin{align}  \label{eq:traveling_wave}
& \bm{m}(x_1,x_2,t) = \bm{m}_0(\xi_1,\xi_2;v_1,v_2) \\
& \xi_1 \equiv x_1-v_1 t,\quad \xi_2=x_2-v_2 t  \notag
\end{align}
so that $\p_t\bm{m} = -v_\nu \p_\nu\bm{m}$ with $\nu=1,2$.
Inserting this in Eq.~\eqref{eq:llg_stt_ip} we obtain
\begin{equation}  \label{eq:llg_stt_ip_steadystate}
\storquedw\,\p_1\bm{m} - v_\nu \p_\nu\bm{m} = 
  -\bm{m}\times \bm{f} 
  + \bm{m}\times \left( \storque\, \storquedw\,\p_1\bm{m} - \alpha v_\nu\, \p_\nu\bm{m} \right).
\end{equation}
We will assume that $\p_1, \p_2$ in the above equation denote derivatives with respect to $\xi_1, \xi_2$.

We may now take the cross product of both sides in Eq.~\eqref{eq:llg_stt_ip_steadystate} with $\p_\mu\bm{m}$ for $\mu=1$ or $2$
and then contract with $\bm{m}$.
In the result, the term containing the effective field $\bm{f}$ (i.e., the term due to the conservative part of the equation)
is written as a total derivative \cite{PapanicolaouTomaras_NPB1991,KomineasPapanicolaou_PhysD1996}:
\begin{equation}  \label{eq:sigma_divergence}
-\bm{f}\cdot \p_\mu \bm{m} = \p_\nu \sigma_{\mu\nu}
\end{equation}
where the explicit form of the tensor $\sigma_{\mu\nu}$ for the effective field \eqref{eq:effective_field}
is given in Ref.~\cite{KomineasPapanicolaou_PRB2015}.
Upon integrating over all space the total derivative \eqref{eq:sigma_divergence} vanishes and we obtain the pair of virial relations \cite{HeLiZhang_PRB2006,EverschorGarst_PRB2011}
\begin{equation}  \label{eq:virial}
\begin{split}
 & ( -4\pi\Skyrmion + \alpha d_{21} ) v_1 + \alpha d_{22} v_2 = \storque\storquedw\, d_{21} - \storquedw\, (4\pi\Skyrmion) \\
 & \alpha d_{11} v_1 + (4\pi\Skyrmion + \alpha d_{12} ) v_2  = \storque\storquedw\, d_{11}.
\end{split}
\end{equation}

Some general conclusions can be drawn from Eqs.~\eqref{eq:virial1_stt_inplane} and \eqref{eq:virial}.
The dynamics and the steady-state velocity $(v_1,v_2)$ can be obtained in particular cases 
as will be discussed in the next sections for the cases of a skyrmionium and a skyrmion.

%%%%%%%%%%
\section{A traveling $\Skyrmion=0$ skyrmionium}
\label{sec:traveling_skyrmionium}

Let us consider the $\Skyrmion=0$ skyrmionium of Fig.~\ref{fig:skyrmionium_vecs} which is a static solution 
of Eq.~\eqref{eq:LL}
and we suddenly apply a spin polarized current.
In order to follow how the initial skyrmionium will be accelerated we will follow 
its {\it linear momentum} which is defined via the conserved $I_\mu$ of Eq.~\eqref{eq:topological_moments} as 
\cite{PapanicolaouTomaras_NPB1991,KomineasPapanicolaou_PhysD1996}
\begin{equation} \label{eq:linear_momentum}
P_\mu = \emn I_\nu,\qquad \mu, \nu = 1\; \hbox{or}\; 2.
\end{equation}
To be sure, the above quantities have the meaning of a linear momentum within the hamiltonian
equations \eqref{eq:LL}. We will though extend their use in the full model \eqref{eq:llg_stt_ip}.
The time derivatives of the components of the linear momentum are given by Eqs.~\eqref{eq:virial1_stt_inplane}
applied for $\Skyrmion=0$:
\begin{equation}  \label{eq:virial1_stt_inplane_Q0}
\begin{split}
(1+\alpha^2) \dot{P}_1 & = (\storque-\alpha)\storquedw\, d_{11} - \alpha\, D_1  \\
(1+\alpha^2)\dot{P}_2 & = (\storque-\alpha)\storquedw\, d_{12} - \alpha\, D_2.
\end{split}
\end{equation}
These relations give a simple result when we apply them for the initial
skyrmionium which is a static solution of Eq.~\eqref{eq:LL} (thus $D_\mu=0$)
and is axially symmetric (thus $d_{12}=0$). We obtain
\begin{equation}  \label{eq:virial1_stt_inplane_Q0_t0}
\dot{P}_1 = \frac{\storque-\alpha}{1+\alpha^2}\storquedw\, d_{11}  \qquad
\dot{P}_2 = 0.
\end{equation}
The skyrmionium is accelerated acquiring a linear momentum component along the $x_1$-direction only.
The acceleration is zero when $\storque = \alpha$ and this point will be clarified in the
following when we present solutions of Eq.~\eqref{eq:llg_stt_ip}.

If the skyrmionium eventually reaches a traveling steady state then 
the virial relations \eqref{eq:virial} apply. For $\Skyrmion=0$ they reduce to a simple form
and give the velocity of the steady state:
\begin{equation}  \label{eq:virial_Q0}
 \begin{pmatrix} d_{11} & d_{12} \\ d_{21} & d_{22} \end{pmatrix}
 \begin{pmatrix} \alpha  v_1 - \storque\storquedw  \\ \alpha v_2  \end{pmatrix}
  = \begin{pmatrix} 0  \\ 0  \end{pmatrix}
\Rightarrow
\begin{cases} v_1 & = \frac{\storque}{\alpha}\storquedw \\ v_2 & = 0 \end{cases}
\end{equation}
provided $\det (d_{\mu\nu}) \neq 0$.
Therefore, the skyrmionium in steady state moves in the direction of the current.
The presence of dissipation is crucial as there is apparently no steady state for $\alpha=0$.
The acceleration process along the axis of the current described by Eqs.~\eqref{eq:virial1_stt_inplane_Q0_t0}
is compatible with the steady-state velocity in Eq.~\eqref{eq:virial_Q0}.

We can find exact traveling solutions for a skyrmionium under spin-transfer torque.
In order to show how this can be achieved, we start by considering the special case
$\storque=\alpha$, for which Eq.~\eqref{eq:llg_stt_ip} becomes
\begin{equation}  \label{eq:llg_stt_ip_special}
(\p_t + \storquedw\, \p_1)\bm{m} = -\bm{m}\times\bm{f} + \alpha\,\bm{m}\times \left( \p_t + \storquedw\, \p_1 \right) \bm{m}.
\end{equation}
We look for traveling wave solutions of the form \eqref{eq:traveling_wave}
and we choose $v_1=\storquedw,\, v_2 = 0$. We have $\p_t\bm{m} = -u\p_1\bm{m}$,
 which is used to reduce Eq.~\eqref{eq:llg_stt_ip_special}
to $\bm{m}\times\bm{f}=0$.
Thus, if we choose $\bm{n}(x_1,x_2)$ to be a static solution of Eq.~\eqref{eq:LL} then
$\bm{m}(x_1,x_2,t)=\bm{n}(\xi_1,x_2)$ gives a configuration which  satisfies Eq.~\eqref{eq:llg_stt_ip_special}
and is a traveling solution with velocity $(v_1, v_2)=(\storquedw,0)$.
In conclusion, the static skyrmionium solution of the conservative LL Eq.~\eqref{eq:LL}
(shown in Fig.~\ref{fig:skyrmionium_vecs})
is a traveling solution of the full equation \eqref{eq:llg_stt_ip} for the case $\storque=\alpha$.
That also explains the vanishing acceleration in Eqs.~\eqref{eq:virial1_stt_inplane_Q0_t0}.
This mathematical result has the following physical content. If we apply spin polarized current to an 
initially static skyrmionium this is expected to be set into rigid motion without significant deformations.
The same can also be argued for the traveling skyrmion in the next Section.

Let us now generalize the above for $\storque \neq \alpha$.
The traveling wave ansatz \eqref{eq:traveling_wave} with the choice $(v_1,v_2)=(v,0)$
is used to reduce Eq.~\eqref{eq:llg_stt_ip} to
\begin{equation}  \label{eq:llg_stt_ip_traveling}
(\storquedw - v)\p_1\bm{m} = -\bm{m}\times\bm{f} + (\storque \storquedw - \alpha v)\,\bm{m}\times \p_1 \bm{m}.
\end{equation}
We now choose $v=\storque\storquedw/\alpha$ and the equation is further reduced to
\begin{equation}  \label{eq:LL_steadystate}
v_0\,\p_1\bm{m} = \bm{m}\times\bm{f},\qquad v_0 = \frac{\storque-\alpha}{\alpha}\,\storquedw.
\end{equation}
This equation is identical to that for a steady-state traveling with a velocity $v_0$
within the conservative LL Eq.~\eqref{eq:LL}.
Such states were numerically calculated and studied in Ref.~\cite{KomineasPapanicolaou_PRB2015}.
A family of traveling skyrmioniums were found with velocities up to a critical velocity $v_c \approx 0.102$.
In conclusion, if we denote the solutions of Eq.~\eqref{eq:LL_steadystate} 
by $\bm{n}(x_1,x_2;v_0)$
then the form $\bm{m}(x_1,x_2,t)=\bm{n}(\xi_1,x_2;v_0)$ with $\xi_1=x_1-vt$ is a 
traveling wave solution of Eq.~\eqref{eq:llg_stt_ip} with velocity $v=\storque\storquedw/\alpha$.
The condition $v_0 < v_c$ for the skyrmionium configuration satisfying Eq.~\eqref{eq:LL_steadystate},
becomes in the presence of spin-transfer torque
\begin{equation}  \label{eq:velocity_condition}
v < u + v_c.
\end{equation}

\begin{figure}[t]
\begin{center}
\includegraphics[width=6truecm]{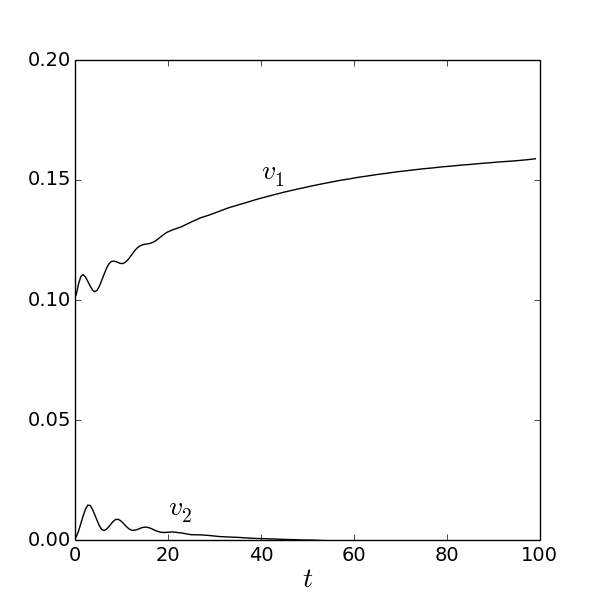}
\caption{The velocity components $(v_1, v_2)=(\dot{\xmagn}_1,\dot{\xmagn}_2)$ for a skyrmionium under spin torque with
parameter values \eqref{eq:parameter_set0}. The expected velocity at the steady state is
$(v_1, v_2) = (0.167, 0)$.}
\label{fig:Q0_velocity}
\end{center}
\end{figure}

We have conducted a numerical simulation 
%on a $400\times 200$ grid 
using the parameter set
\begin{equation}  \label{eq:parameter_set0}
\alpha = 0.06,\qquad \storquedw = 0.1,\qquad \storque = 0.1.
\end{equation}
We use as an initial condition the skyrmionium of Fig.~\ref{fig:skyrmionium_vecs} with $\anisotropy=3$
and apply the spin current.
The expected velocity in the steady-state is given by Eq.~\eqref{eq:virial_Q0}
and is $v_1=0.167,\; v_2=0$.
The prediction is confirmed by the numerical simulation.
We define the position of the skyrmionium as $(\xmagn_1,\xmagn_2)$ with
\begin{equation}  \label{eq:moments_magnetization}
\xmagn_\mu = \frac{\int x_\mu(1-m_3)\, d^2x}{\int (1-m_3)\, d^2x}.
\end{equation}
Fig.~\ref{fig:Q0_velocity} shows the velocity $(v_1, v_2)=(\dot{\xmagn}_1,\dot{\xmagn}_2)$ as a function of time.
The skyrmionium is accelerated and the change of linear momentum at $t=0$ verifies
the prediction $\dot{P}_1 = 0.34$ calculated by Eq.~\eqref{eq:virial1_stt_inplane_Q0_t0} 
when we use the numerically calculated value $d_{11}=57.1$.
The velocity at $t=0$ is $v_1(t=0) \approx \storquedw = 0.1$ and it increases to $v_1(t=100) = 0.16$
at the end of this simulation.
The component $v_2$ acquires some small value at the initial stages of the simulation and it later goes to zero.
The results have been confirmed also by a simulation in a moving frame,
running for times longer than those shown Fig.~\ref{fig:Q0_velocity}, 
where it is seen that $(v_1,v_2)$ converge to the expected values at the steady-state.

\begin{figure}[t]
\begin{center}
\includegraphics[width=4truecm]{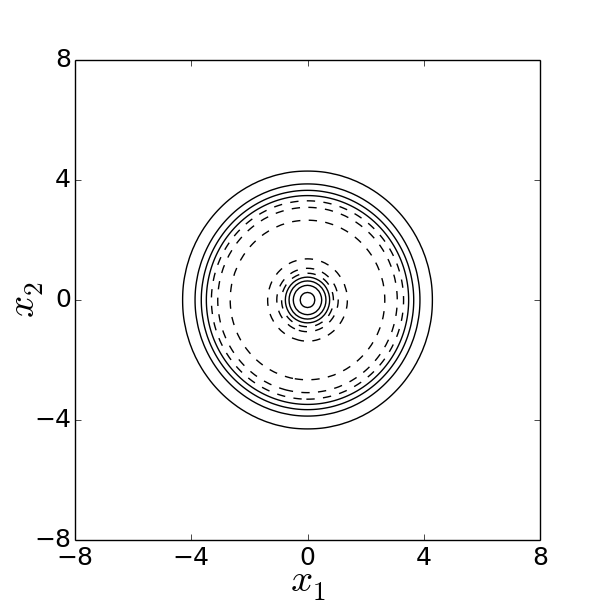}
\includegraphics[width=4truecm]{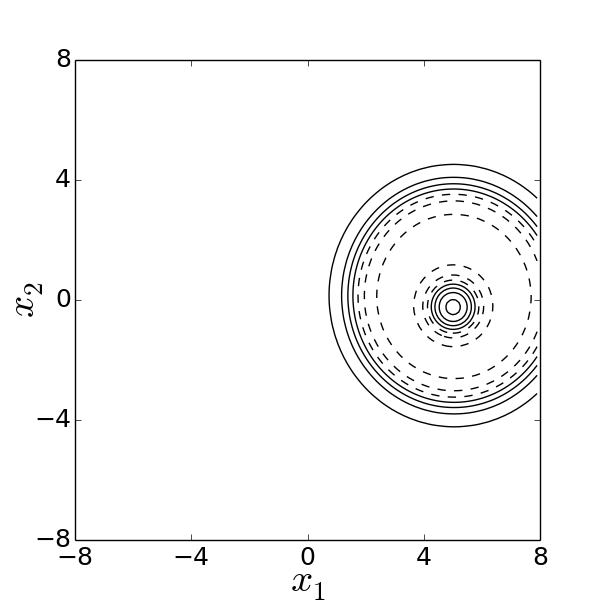}
\caption{Contour plots of $m_3$ for a skyrmionium under spin torque with
parameter values \eqref{eq:parameter_set0}.
(Left) The initial condition, at $t=0$, is a static skyrmionium solution of Eq.~\eqref{eq:LL}.
(Right) The skyrmionium at $t=40$ when it has been accelerated.
The inner part has moved down relative to the outer part.
The contour levels plotted are $m_3=0.9,0.6,0.3,0.0$ (solid lines) and $m_3=-0.3,-0.6,-0.9$ (dashed lines).
}
\label{fig:Q0_stt}
\end{center}
\end{figure}

Fig.~\ref{fig:Q0_stt} shows two snapshots of the skyrmionium under spin torque.
At the initial time $t=0$ we have the axially symmetric static solution of Eq.~\eqref{eq:LL}.
The accelerated skyrmionium at $t=40$ has velocity $v_1 = 0.143$ and has lost axial symmetry:
its central part has moved lower.
It is very similar to the propagating skyrmionium studied in Ref.~\cite{KomineasPapanicolaou_PRB2015}.
We conclude that the application of spin current is a method to obtain a propagating skyrmionium
in a steady state.
We note that the skyrmionium continues to travel at its acquired velocity when the spin currents
is switched off, irrespectively of whether a steady state was reached or not.
This is in stark contrast to the dynamics of a skyrmion or to ordinary domain wall dynamics.

Let us now consider a second set of parameter values
\begin{equation}  \label{eq:parameter_set1}
\alpha = 0.04,\qquad \storquedw = 0.1,\qquad \storque = 0.1.
\end{equation}
which gives a velocity for the skyrmionium $v=\storque\storquedw/\alpha=0.25$
violating condition \eqref{eq:velocity_condition}.
In this case the skyrmionium is accelarated until its velocity approaches
the limiting value $\storquedw + v_c$ while the configuration becomes elongated along the $x_2$ axis and eventually
reaches the boundaries of our numerical grid.
Presumably, the process would continue until the skyrmionium configuratiom is destroyed
or until it turns to a domain wall extending to infinity in the $x_2$ direction.

%%%%%%%%%%%%%%%%%%%%%%%%%%%%%%%%%%%%%%%%
\section{A traveling $\Skyrmion=1$ skyrmion}
\label{sec:traveling_skyrmion}

Let us now consider topologically nontrivial solutions ($\Skyrmion \neq 0$)
such as the $\Skyrmion=1$ skyrmion of Fig.~\ref{fig:skyrmion_vecs} which is a static solution 
of Eq.~\eqref{eq:LL},
and suddenly apply a spin polarized current.
In order to follow the skyrmion as it moves we will follow the coordinates
of its guiding center $(R_1, R_2)$
defined as the normalized moments in Eq.~\eqref{eq:topological_moments}:
\begin{equation}  \label{eq:guiding_center}
R_\mu = \frac{I_\mu}{4\pi\Skyrmion} = \frac{1}{4\pi\Skyrmion} \int x_\mu\skyrmion\, d^2x.
\end{equation}
They give a measure of the position of a skyrmion and are conserved quantities
as explained in connection with Eqs.~\eqref{eq:topological_moments}, \eqref{eq:vorticity_time_derivative}.

The instantaneous velocity $(\dot{R}_1, \dot{R}_2)$ is given through Eqs.~\eqref{eq:virial1_stt_inplane}.
For the initial axially symmetric skyrmion solution of Eq.~\eqref{eq:LL}
we have $D_\mu=0$ and $d_{12}=0,\; d_{11}=d_{22}= \Eex$, where $\Eex$ is the exchange energy.
Eq.~\eqref{eq:virial1_stt_inplane} gives
\begin{equation}  \label{eq:Rdot_static}
\dot{R}_1 = \storquedw \frac{1+\alpha\storque}{1+\alpha^2} = \storquedw + \frac{\alpha}{\bar{d}}\dot{R}_2\,,  \qquad
\dot{R}_2 = \storquedw\frac{(\storque-\alpha)\, \bar{d}}{1+\alpha^2}
\end{equation}
where we denoted $\bar{d}=d_{11}/(4\pi\Skyrmion)=d_{22}/(4\pi\Skyrmion)$.
Thus, the skyrmion will initially have a velocity component in the direction of the current flow while 
its velocity component perpendicular to it depends on the sign of $\storque-\alpha$.

If we now assume that the skyrmion will eventually reach a propagating steady state
then its velocity will satisfy the virial relations \eqref{eq:virial}.
For $\Skyrmion \neq 0$ we define $\bar{d}_{\mu\nu} = d_{\mu\nu}/(4\pi\Skyrmion)$ 
and write the virial relations as
\begin{equation}  \label{eq:virial_velocity_topological}
\begin{split}
 & ( 1 - \alpha \bar{d}_{12} ) v_1 - \alpha \bar{d}_{22} v_2 = (1 - \storque\bar{d}_{12})\storquedw \\
 & \alpha \bar{d}_{11} v_1 + (1 + \alpha \bar{d}_{12} ) v_2  = \storque\storquedw\, \bar{d}_{11}.
\end{split}
\end{equation}
These imply that, in general, both components of the velocity are nonzero, $v_1, v_2 \neq 0$,
therefore a propagating skyrmion will move at an angle to the flow of the spin current.
Unlike in the case of a skyrmionium, 
dissipation is not necessary in order to obtain a steady-state in the case of the skyrmion and 
Eqs.~\eqref{eq:virial_velocity_topological} give for $\alpha=0$ the velocity
\begin{equation}
v_1=\storquedw - \storque\storquedw\,\bar{d}_{12},\qquad v_2= \storque\storquedw\,\bar{d}_{11}.
\end{equation}

In order to obtain more detailed information on the propagating skyrmion configuration
we study the case $\storque=\alpha$ since it emerges again as a special case as seen in 
Eq.~\eqref{eq:Rdot_static}.
In a steady state Eq.~\eqref{eq:virial_velocity_topological} gives for
the skyrmion a velocity $(v_1,v_2)=(\storquedw,0)$ colinear with the spin current.
When we substitute this in Eq.~\eqref{eq:llg_stt_ip_steadystate}  
we obtain the static LL equation
$\bm{m}\times\bm{f}=0$. Thus the argument employed in Sec.~\ref{sec:traveling_skyrmionium}
for a skyrmionium also applies for a skyrmion:
a static skyrmion solution of the conservative LL Eq.~\eqref{eq:LL}, which we denote $\bm{n}(x_1,x_2)$,
is a traveling solution of the full equation \eqref{eq:llg_stt_ip} 
with $\bm{m}(x_1,x_2,t)=\bm{n}(\xi_1,x_2)$ and $\xi_1=x_1-\storquedw t$.
 
It is well-known that there are no traveling skyrmion solutions of Eq.~\eqref{eq:LL},
i.e., there are no $\Skyrmion \neq 0$ skyrmion solutions of Eq.~\eqref{eq:LL_steadystate},
and this can be rigorously established \cite{PapanicolaouTomaras_NPB1991}.
Therefore, the arguments about traveling skyrmionium solutions following Eq.~\eqref{eq:LL_steadystate}
cannot be applied in the case of skyrmions.

For $\storque\neq \alpha$ a traveling skyrmion should have $v_2 \neq 0$ as shown
by Eq.~\eqref{eq:virial_velocity_topological}, that is, the skyrmion travels at an angle with respect to
the direction of the flow of the spin current.
%This phenomenon has an origin analogous to the Magnus force in the case of the LL equation
%under external forces.
In the case $\storque \neq \alpha$ we could not find exact traveling wave solutions of Eq.~\eqref{eq:llg_stt_ip},
for the effective field \eqref{eq:effective_field}.
Exact results are though indeed obtained for the
pure exchange model in Appendix \ref{sec:exchange_model}.

For small deviations from the simple case $\storque=\alpha$ and $(v_1,v_2)=(\storquedw,0)$
we may assume that the skyrmion retains approximately axial symmetry,
and thus we have a diagonal $\bar{d}_{\mu\nu}=\bar{d}\,\delta_{\mu\nu}$ where $\bar{d}$ is a constant.
Eqs.~\eqref{eq:virial_velocity_topological} have now a relatively simple solution:
\begin{equation}  \label{eq:virial_velocity_topological_simple}
 v_1 =  u\frac{1 + \alpha\storque \bar{d}^2}{1 + (\alpha \bar{d})^2}
 = \storquedw + \alpha \bar{d}\,v_2,  \quad
 v_2 = u\frac{(\storque-\alpha)\,\bar{d}}{1 + (\alpha \bar{d})^2}.
\end{equation}
For the pure exchange model $\bar{d}=1$ (see Appendix \ref{sec:exchange_model})
while for other models such as in Eq.~\eqref{eq:effective_field} we have $\bar{d}>1$
calculated by substituting the static skyrmion solution of Eq.~\eqref{eq:LL} in Eq.~\eqref{eq:d_munu}.
Eq.~\eqref{eq:virial_velocity_topological_simple} gives the 
so-called mobility relation, i.e., a linear relation between the velocity and the current.

\begin{figure}[t]
\begin{center}
\includegraphics[width=6truecm]{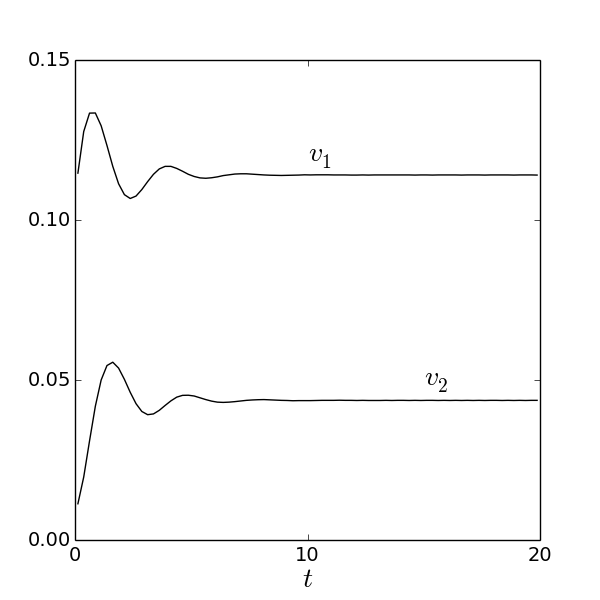}
\caption{Simulation results for the velocity components $(v_1, v_2)=\dot{R}_1, \dot{R}_2)$ as a function of time $t$ 
for a skyrmion under spin torque with parameter values \eqref{eq:parameter_set1b}.}
\label{fig:Q1_velocity}
\end{center}
\end{figure}

In order to check the theoretical predictions
we have conducted a numerical simulation using the parameter set
\begin{equation}  \label{eq:parameter_set1b}
\alpha = 0.2,\qquad \storquedw = 0.1,\qquad \storque = 0.5
\end{equation}
with large values for $\alpha,\storque$ and $\storque-\alpha$.
We use as an initial condition the skyrmion of Fig.~\ref{fig:skyrmion_vecs} with $\anisotropy=3$
and apply the spin current.
The numerically found velocity for the skyrmion is shown in Fig.~\ref{fig:Q1_velocity}.
We initially have $\dot{R}_1 = 0.106,\; \dot{R}_2 = 0.044$; 
the velocity presents oscillations and eventually converges to constant values $(v_1, v_2) = (0.1140, 0.0436)$.
The expected initial velocity is found from Eqs.~\eqref{eq:Rdot_static} and it is
$\dot{R}_1 = 0.106,\; \dot{R}_2 = 0.046$, where we have used the numerically calculated value
for the exchange energy $\Eex=20.08 \Rightarrow \bar{d} = 1.598$.
The final velocity is in excellent agreement with the velocity at a steady state predicted
by Eq.~\eqref{eq:virial_velocity_topological_simple} which gives $(v_1, v_2) = (0.1139, 0.0435)$,
where we assume that the initial skyrmion profile is not significantly changed.
The skyrmion configuration is indeed not visibly distorted during the simulation
compared to the initial axially symmetric skyrmion.

We have also conducted a numerical simulation using the parameter set
\eqref{eq:parameter_set1} and the results are again in excellent agreement with the predictions
of Eqs.~\eqref{eq:Rdot_static} and \eqref{eq:virial_velocity_topological_simple}.

We finally note that the skyrmion is pinned at its final position when the spin current is switched off.
This is contrasted with the dynamics of a skyrmionium which travels at a constant velocity 
also in the absence of external forces (and damping).

%%%%%%%%%%%%%%%%%%%%%%%%%%%%%%%%%%%%%%%%
\section{Concluding remarks}
\label{sec:conclusions}

We have studied the dynamics under spin torque 
of non-topological $(\Skyrmion = 0)$ and topological $(\Skyrmion \neq 0)$ skyrmions
in films of Dzyaloshinskii-Moriya materials with easy-axis anisotropy.
Analytical results are obtained using the equations for the linear momentum (for $\Skyrmion = 0$)
or the guiding centre (for $\Skyrmion \neq 0$), and virial relations.
The study is complemented by a set of numerical simulations.
Furthermore, we obtain exact solutions of the Landau-Lifshitz equation including spin torques,
given in Eq.~\eqref{eq:llg_stt_ip}, for some particular cases.
The traveling solutions obtained provide arguments that a spin polarized current will cause
rigid motion of a skyrmion or a skyrmionium.
This result is applicable not only for the solitons studied in the present work but also
for other models, e.g., for driven motion of magnetic bubbles.

The $\Skyrmion=0$ skyrmionium is accelerated by the spin torque and it continues moving
after switching off the current. This Newtonian dynamics was also observed in the case
of the skyrmionium under an external field gradient \cite{KomineasPapanicolaou_PRB2015}.
It is dramatically different than the more well-studied dynamics of a skyrmion
\cite{FertCros_NatNano2013,IwasakiMochizuki_NatComms2013,IwasakiMochizuki_NatNano2013} which is spontaneously pinned in the absence of external torques.
On the other hand, the skyrmionium motion presents some similarities with the skyrmion motion in a stripe geometry
\cite{IwasakiMochizuki_NatNano2013}.

Our methods can be also applied to simpler one-dimensional models (wires) provided static domain wall
solutions of Eq.~\eqref{eq:LL} exist.

\appendix
%%%%%%%%%%%%%%%%%%%%%%%%%%%%%%%%%%%%%%%%
\section{The pure exchange model}
\label{sec:exchange_model}

If we set $\bm{f}=\Delta\bm{m}$ we have the so-called pure exchange model
and we will present analytic results which elucidate the discussion in Sec.~\ref{sec:traveling_skyrmion}.
In the pure exchange model the Bogomolnyi relations \cite{BelavinPolyakov_JETP1975,Rajaraman}
\begin{equation}  \label{eq:Bogomolnyi_BP}
\p_1\bm{m} = \bm{m}\times\p_2\bm{m},\qquad \p_2\bm{m} = -\bm{m}\times\p_1\bm{m},
\end{equation}
which contain only first order derivatives,
are sufficient for obtaining static solutions of Eq.~\eqref{eq:LL}, 
i.e., solutions for $\bm{m}\times\bm{f}=0$.
A large class of $\Skyrmion \neq 0$ skyrmion solutions can be found by solving \eqref{eq:Bogomolnyi_BP}.
Of those, the axially symmetric $\Skyrmion = 1$ skyrmion configuration of the form \eqref{eq:axially_symmetric} 
will be denoted $\bm{m}=\bm{n}(x_1,x_2)$ and reads
\begin{equation}  \label{eq:axially_symmetric_Q1}
n_1 = -\frac{2a x_2}{\rho^2+a^2},\quad
n_2 = \frac{2a x_1}{\rho^2+a^2},\quad
n_3 = \frac{\rho^2-a^2}{\rho^2+a^2},
\end{equation}
where $a$ is a arbitrary positive constant giving the skyrmion radius and $\rho^2=x_1^2+x_2^2$.
Configuration \eqref{eq:axially_symmetric_Q1} is similar in its gross features to that shown in Fig.~\ref{fig:skyrmion_vecs}.

We turn to Eq.~\eqref{eq:llg_stt_ip_steadystate} for a traveling steady state
with velocity $(v_1, v_2)$ and we require
\begin{equation}  \label{eq:velocity_topological_bogomolny}
\begin{cases}
& \storquedw - v_1 = -\alpha v_2 \\
& v_2 = \storque\storquedw - \alpha v_1
\end{cases}
\Rightarrow
\begin{cases}
 v_1 & = \frac{1+\alpha\storque}{1+\alpha^2}\, \storquedw \\
 v_2 & = \frac{\storque-\alpha}{1+\alpha^2}\, \storquedw.
\end{cases}
\end{equation}
Then, Eq.~\eqref{eq:llg_stt_ip_steadystate} simplifies to
\begin{equation}  \label{eq:llg_stt_ip_steadystate_exchangemodel}
\left( \alpha v_2\,\p_1 + v_2 \p_2 \right)\bm{m} =
\bm{m}\times\bm{f} + \bm{m}\times \left( \alpha v_2\p_2 - v_2\,\p_1 \right) \bm{m}.
\end{equation}
Under the Bogomolnyi relations \eqref{eq:Bogomolnyi_BP}
the terms on the left-hand-side cancel with the last terms on the right-hand-side 
which originate in the damping and the non-adiabatic spin torque.
Since the same Bogomolnyi relations are sufficient for the vanishing of the first term on the right-hand-side,
we conclude that relations \eqref{eq:Bogomolnyi_BP} are sufficient conditions for solutions
of Eq.~\eqref{eq:llg_stt_ip_steadystate_exchangemodel}.

Therefore, any solution $\bm{m}=\bm{n}(x_1,x_2)$ of Eqs.~\eqref{eq:Bogomolnyi_BP},
such as the skyrmion in Eq.~\eqref{eq:axially_symmetric_Q1}, 
is a traveling solution $\bm{m}(x_1,x_2,t)=\bm{n}(\xi_1,\xi_2)$ of the full Eq.~\eqref{eq:llg_stt_ip} 
with spin current and damping, with a velocity given by Eq.~\eqref{eq:velocity_topological_bogomolny}.

Regarding calculations presented in Sec.~\ref{sec:traveling_skyrmion} it is useful to note that
skyrmion solutions with $\Skyrmion \neq 0$
which satisfy \eqref{eq:Bogomolnyi_BP} have exchange energy $\Eex=4\pi\Skyrmion$ \cite{Rajaraman}.
Thus, for axially symmetric skyrmions, such as \eqref{eq:axially_symmetric_Q1},
the tensor $d_{\mu\nu}$, defined in Eq.~\eqref{eq:d_munu}, is diagonal with $\bar{d}_{\mu\nu} = \delta_{\mu\nu}$.
We then see that the velocity in Eq.~\eqref{eq:velocity_topological_bogomolny} 
coincides with that given in Eq.~\eqref{eq:virial_velocity_topological_simple} applied for $\bar{d}=1$.

%\acknowledgements

\end{document}